# Machine Learning Reveals the State of Intermittent Frictional Dynamics in a Sheared Granular Fault

**Authors:**

C. X. Ren[1,2†*], O. Dorostkar[3†*],

B. Rouet-Leduc[2], C. Hulbert[2], D. Strebel[4], R. A. Guyer[2], P. A. Johnson[2], J. Carmeliet[3]

[1] Space Data Science and Systems Group, Los Alamos National Laboratory, MS D440, Los Alamos, New Mexico 87545, USA

[2] Geophysics Group, Los Alamos National Laboratory, MS D446, Los Alamos, New Mexico 87545, USA

[3] Department of Mechanical and Process Engineering, Swiss Federal Institute of Technology Zürich (ETH Zürich), Tannenstrasse 3, CH-8006 Zürich, Switzerland

[4] Swiss Federal Laboratories for Materials Science and Technology (Empa), Überlandstrasse 129, CH-8600 Dübendorf, Switzerland

†Equal contribution
*Corresponding authors:

Omid Dorostkar        (Email: domid@ethz.ch) (ID: orcid.org/0000-0002-7758-4919)
Christopher. X. Ren    (Email: cren@lanl.gov) (ID: orcid.org/0000-0002-0787-6713)

**Abstract**

The seismogenic plate boundaries are presumed to behave in a similar manner to a densely packed granular medium, where fault and blocks systems rapidly rearrange the distribution of forces within themselves, as particles do in slowly sheared granular systems. We use machine learning and show that statistical features of velocity signals from individual particles in a simulated sheared granular fault contain information regarding the instantaneous global state of intermittent frictional stick-slip dynamics. We demonstrate that combining features built from the signals of more particles can improve the accuracy of the global model, and discuss the physical basis behind decrease in error. We show that the statistical features such as median and higher moments of the signals that represent the particle displacement in the direction of shearing are among the best predictive features. Our work provides novel insights into the applications of machine learning in studying frictional processes that take place in geophysical systems.

# 1 Introduction

Characterizing the state of friction in granular flows in Earth can significantly improve our understanding of geological and geophysical frictional processes that take place in earthquakes, landslides, avalanches, debris flows and soil liquefaction [van der Elst et al., 2012; Brzinski & Daniels, 2018; Rouet-Leduc et al., 2018]. In a critical loading configuration, sheared granular layers can exhibit stick-slip dynamics that resemble the intermittent dynamics of earthquakes [Brace & Byerlee, 1966, 1970; Johnson et al., 1973; Scholz, 1998]: during the stick phase, which is the earthquake nucleation phase, the elastic strain energy builds up and at slip, the energy is released as an earthquake [Dorostkar & Carmeliet, 2018].

Sheared granular systems can be considered analogous to the macroscopic dynamics of faults in Earth [Anderson, 2007], where both systems exhibit intermittent dynamics: motion at boundaries results from the differential motion of the particles or juxtaposed tectonic blocks; the release of energy is due to rearrangement of particles, in sheared granular layers or earthquakes in fault zones [Meroz & Meade, 2017]. The statistical distributions of slip events in both laboratory and numerical fault gouge are shown to follow the Gutenberg-Richter law, providing similar b-values [Gutenberg & Richter, 1944; Dahmen et al., 2011; Uhl et al., 2015; Dorostkar, 2018; Rivière et al., 2018]. The analysis of fluctuations in Global Positioning System (GPS) observations of inter-seismic motion from the southern California plate boundary has identified the same statistical distribution of velocity fluctuations as in slowly sheared granular media [Meroz & Meade, 2017]. This may suggest that the plate boundary can be understood as a densely packed granular medium, predicting a characteristic tectonic length scale and relating the characteristic duration and recurrence interval of earthquakes [Meroz & Meade, 2017].

Numerical tools such as Discrete Element Method (DEM) simulations are a useful means to shed light on the frictional processes that govern fault slip [Dorostkar, Johnson, et al., 2017; Dorostkar et al., 2018]. Numerical simulations can provide micro-scale information [Dorostkar, Guyer, et al., 2017] whilst most experimental studies of stick-slip dynamics are opaque. Further, assessing the internal stress state of a granular system is notoriously difficult, and even photo-elastic, optical, and tomographic techniques require specialized materials or slow scanning times [Brzinski & Daniels, 2018]. DEM simulations can provide insight into the inner workings of such systems clarifying the complexities of frictional behaviour. Furthermore, DEM modelling enables us to extract the information needed to build models for characterizing the frictional state of a fault system subject to shear.

The recent applications of machine learning in earthquake identification [Perol et al., 2018; Ross et al., 2018], association for seismic arrivals [McBrearty et al., 2019], estimation of fault displacement rate in the Cascadia subduction zone [Rouet-Leduc et al., 2019], detection of similarities between slow and fast slip events [Hulbert et al., 2019] and prediction of remaining time to failure [Rouet-Leduc et al., 2017; Corbi et al., 2019] are promising developments underscoring the rich potential of modern data analysis techniques. Machine learning techniques can help us to better understand and address complex open questions in geoscience, since they can augment our intuition and help to reveal structure in high-dimensional datasets [Kong et al., 2018; Ross et al., 2019].

Capitalizing on the advantages of DEM simulations and machine learning techniques, here we use a distributed gradient boosting algorithm [J. Friedman et al., 2000; J. H. Friedman, 2002; Chen & Guestrin, 2016] to build models that are able to estimate the highly irregular frictional state in a sheared granular media. By using the velocity signals of individual flagged particles inside the granular fault, the Extreme Gradient Boosting (XGBoost) model in this paper shows that the individual grains contain predictive information concerning the global frictional state during a stick-slip cycle. We posit that based on the statistical similarities between our numerical simulations, laboratory experiments and plate boundary scale GPS velocity fluctuations [Meroz & Meade, 2017; Rouet-Leduc et al., 2019], the granular approximation in numerical simulations provides a useful mathematical framework for understanding and characterizing the frictional processes that take place in a tectonic plate boundary [Meroz & Meade, 2017]. Our developments and findings in this paper provide novel insights into the dynamics of sheared granular systems, which we believe open promising windows for future applications of machine learning in studying frictional processes that take place in geophysical systems.

## 2 Materials and Methods

Figure 1a illustrates the DEM granular layer with size of 11×1.5×0.8 $mm^3$ containing 7996 spherical particles with a uniform, poly-disperse particle size distribution ranging 90-150 μm. On the sample top and bottom, we use two corrugated plates with high surface roughness modelled by a friction coefficient of 0.9 between the plates and particles to totally engage the fault blocks with granular particles (see Fig.1a insets). The position of the upper corrugated plate is adapted continuously to maintain constant the confining stress. A displacement-controlled servomechanism moves the bottom corrugated plate along the x direction at constant velocity of 600 μm/s. The confining stress is 10 MPa. On the front- and back-side of the sample, we employ frictionless walls with the same elastic properties as the particles to avoid rigid wall boundary conditions. Periodic boundary conditions are applied at the left and right sidewalls in x direction. The particle density is 2900 $kg/m^3$, which results in a DEM time step of $15\times10^{-9}$ seconds. The granular flow remains in the quasi-static regime by setting the inertial number to be below $10^{-3}$ [MiDi, 2004]. We use LIGGGHTS [Goniva et al., 2012; Kloss et al., 2012] to model the granular system. The properties of DEM model is fully presented in Table 1 of Supplementary Information.

We test for the relationship between the instantaneous global frictional state of the simulated system and the particle velocity by applying a supervised learning approach. We calculate statistical features of the particle velocity signal for short moving time windows, as described in [Rouet-Leduc et al., 2017]. These pre-determined features consist of various higher-order statistical moments, percentiles, and inter-quartile ranges within the windows. The machine learning (ML) model utilizes these features in order to estimate the instantaneous, global friction coefficient of the system for the final step of a time window for which the features are extracted. The time windows considered in this work consist of 10 time steps from the DEM simulation, equal to $15\times10^{-8}$ seconds, and overlap each other by 9 time steps. Finally, we smooth the features across 30 time steps to achieve the best results.

We utilize the XGBoost implementation [Chen & Guestrin, 2016] of an ML algorithm known as gradient boosted decision trees [J. H. Friedman, 2002], an ensemble method for decision trees [Breiman, 1984], with an L2 loss function (minimizing mean squared error). We choose L2 over L1 as a loss function, as in this case we are looking to capture the behavior of the slip events, which can be considered 'outliers' in the overall distribution of the macroscopic friction values of the system. Boosting is a strategy whereby multiple 'weak' models are sequentially combined into a 'strong' composite model by fitting each new weak learner to the residuals of the previous iteration (see Supplementary Information).

The training set for this model comprises the first 80% of the simulated data and the testing set used to evaluate the performance of the model consists of the remaining 20%. We train the gradient boosting model by providing the features derived from the sliding time windows as input, and the macroscopic friction calculated from the DEM simulations as label. We then test the trained model on the testing set, by providing only the features of the velocity signal from a single particle attempting to estimate the unseen friction of the system during this period. The performance of our model is then evaluated with respect to the test ground truth (macroscopic friction calculated from the DEM simulations as label) using the coefficient of determination as an evaluation metric.

In order to evaluate the relative importance of features, we utilize Shapely Additive Explanations (SHAP) values, a relatively novel method of attributing feature importance based on game theory [S. M. Lundberg & Lee, 2017]. Traditionally, global feature importance for decision tree algorithms are calculated using gain, split count, or permutation methods [Auret & Aldrich, 2011]. However, it has been shown that feature importance values calculated in the aforementioned manners are inconsistent: the mechanics of a model can change such that it relies more on a given feature to make predictions, yet the importance estimate of said feature can decrease [Scott M Lundberg et al., 2018]. SHAP values combine Shapley values from game theory [Shapley, 1953] with the conditional expectation function of the model. Given a subset of input features, the features are modeled as 'players' in a co-operative game where the end goal is prediction. The 'pay-outs' from the game are thus the feature importance, and calculated by determining the contribution of each 'player' to the game [S. M. Lundberg & Lee, 2017] (See Supplementary Information for more details).

## 3 Results

For training process, the feature space is constructed for 4 properties of the particles i.e. the x, y and z-components of particle velocity (denoted in this work as $V_x$, $V_y$ and $V_z$) together with its magnitude, V. Fig. 2a shows the time series of macroscopic friction and $V_x$ for particle 2146 on the right-hand y-axis. We observe continuous fluctuations during the stick phase and rapid bursts in particle velocity associated with each slip event. After the training of the ML model (Fig. 2b), the velocity signal of particle id 2146 exhibits the best performance among all of the particles in the system with a model performance of $R^2 = 0.57$ in Fig. 2c. The $R^2$ is calculated for the gradient boosting model trained on statistical features from the velocity signal of a single particle using the process described in Sec. 2. This observation shows that a single particle can be used to determine the macroscopic friction of the entire system it belongs to: the statistical characteristics of its velocity signal are a fingerprint of the global frictional state.

We perform the ML analysis for the velocity signals of all particles in the fault model; Fig. 3a shows the distribution of $R^2$ scores for 7996 particles during the testing period. The median of this distribution is $R^2 = 0.09$, with a 90[th] percentile value of $R^2 = 0.3$, indicating that a large number of particles carry limited useful information concerning the macroscopic frictional state of the system over the testing period. We posit that since a single particle is not involved in the entire structure of granular contacts, and that some of the slip events may be localized in a relatively small region of the system, the particle may not be sensitive to rearrangements far away from its position.

We also study the feature importance in our ML model that is developed from the velocity signal of a single particle. Figure 3b shows the ten most important features for the ML model giving the results shown in Fig. 2. The feature importance is expressed in terms of mean absolute SHAP values, as detailed in Sec. 2. The impact of each individual instance of the features at each row is shown on the x-axis. We note that two complementary features display the largest range of SHAP values: the median value of $V_x$, and the 4$^{th}$ moment of V. These two features are complementary in terms of their statistical properties: the median of a distribution is robust to outliers, whereas the 4$^{th}$ moment is in fact a measure of 'heaviness' of the tail of a distribution, and is extremely sensitive to outliers [Đorić et al., 2009]. We remark that these two most important features are followed narrowly by the 4$^{th}$ moment of $V_y$ (Fig. 3b), indicating that slip events are not sensed by the particle only in the shearing direction, but also in the direction of system compaction at slip [Dorostkar & Carmeliet, 2019]. Interestingly, a moving median provides a robust estimate of the trend of a time series when variations from the trend are Laplace distributed [Arce, 2005], as is the case with the velocity of the particle. The features deemed important by the model can thus be interpreted as the underlying trend of the velocity (or energy) of the particle, as well as an 'event detector' defined by the 4$^{th}$ moment of the velocity. The relationship between the underlying trend of the particle energy and the macroscopic friction of the system is in good agreement with the results reported in laboratory by Rouet-Leduc *et al.* [2018], where the key feature for the estimation of fault friction is the variance of the acoustic signal, which is directly related to the seismic energy contained in a given time window.

Figure 3c shows a SHAP dependence plot, similar in nature to partial dependence plots often used for the interpretation of ML models [Greenwell, 2017]. The feature instances are colored by the values of another feature (in this case, the median of $V_x$) to reveal potential feature interactions. Here we show the logarithm of the 4$^{th}$ moment of x-component of the particle velocity, plotted against SHAP value, or 'importance' as discussed in Sec. 2. This feature clearly dominates the model estimations as it approaches large values, corresponding to larger slip events, as shown in Fig. 3f, where sudden increases in the 4$^{th}$ moment (green trace) align with large drops in the macroscopic friction, confirming that the 4$^{th}$ moment of the particle velocity operates as an event detector in our model

Figure 3d shows the SHAP dependence plot for the median of $V_x$, colored by the logarithm of the 4$^{th}$ moment. We observe two regimes where the median has an impact on the ML model: one where the value of the 4$^{th}$ moment is large (approximately >10$^{-1}$), and one where it is far smaller (~7 orders of magnitude). We speculate this interaction exists due to the fact that large slip events can induce changes in the underlying trend of the velocity of the particle. This is shown in Fig. 3e where the median (red trace) correlates with small variations in the friction between the larger slip events, but also decreases gradually across these events, which are sensed by the 4$^{th}$ moment of the velocity.

We demonstrate that the performance of the ML model can be improved by combining the features used for the single particle mode with features derived from the velocity of a different particle. Figure 4a shows the improved model (black trace) plotted against the best single particle model (red trace) and ground truth (blue trace). Although the model does not always estimate a more accurate value for the macroscopic friction across the test set, the improvements made by including the features from the second particle are substantial: the test $R^2$ of the model improves to a value of 0.7. In particular, we note the model error is vastly reduced towards the end of the test set, in the time steps ranging between 3.3-3.5 × 10$^8$, where the 4$^{th}$ moment of the combined particle exhibits more spikes, corresponding to the response of the particle to slip events occurring in the system, as is shown in Fig. 4b. The increase in performance of the ML model shows that recording the velocity signals in more regions of the fault system may help to estimate the state of friction more accurately. Adding more particles in the preparation of the feature space does not necessarily improve the performance of the model as the model requires a consistent mapping between the training and testing sets.

In Fig. 4c, we show the change in distribution of test $R^2$ scores between the single best particle (ID=2146, shown in red), and the features from this particle (ID = 2146) combined with features built from all other particles in the simulation (black). We hypothesize that the ML model improves as it has access to more information concerning the state of the system, but through the similar features as for the single particle. This fact is evidenced in Fig. 4d, which shows that the top ten most important features for the combined ML model mainly consists of the skewness measures, as well as features consisting of the moving median, which are similar to those found for the single particle, shown in Fig. 3b. Interestingly, of the 7995 particles shown in the combined distribution in Fig. 4c (black), only 189 combinations (approximately 2%) improve the test $R^2$ beyond that of the model built using only features from particle 2146. Based on these results we can conclude that not only can the instantaneous global state of the simulated system be recovered from the characteristics of a single particle in this system, but that this particle is 'privileged' in terms of the information it has access to, as only 2% of the particles provide complementary information when added to the model. We speculate that this 'privileged' position is likely due to the relative position of the particle and the force network through the system: the particle must be somewhat free to move in order to transmit information concerning global state of the system, but also close enough to areas of change in the force network to avoid uninformative 'rattling' behavior. This extremely interesting observation with regards to the information content of the bead pack is something beyond the scope of this letter but will be the subject of future investigations.

## 4 Summary and Conclusions

We simulate a sheared granular layer and use machine learning to estimate the instantaneous, bulk friction of system. We show that the velocities of individual particles contain predictive information regarding the global state of the system. We also show that combining signals of more particles can improve our models' ability to capture the evolution of frictional state of the system. Our analyses show that the statistical features i.e. median and higher moments of signals that represent the particle displacement in the direction of shearing during the stick phase and the particle displacement magnitude at slip event are among the best estimators of the friction of the system.

The fact that only some particles in the system offer instantaneous predictability of the friction is fascinating and demonstrates not all grain velocities are equal in their information content. This may be due to non-stationarity in the simulation, as the training set may not fully contain informative velocity behavior for modelling the testing set. When analyzing the bulk velocity, there may be cancellation effects of the ensemble that diminish instantaneous predictability. We have observed that the most predictive grains are located near regions of high stress force chains, but not in them. The predictive grains are also not 'rattlers'. The relative contributions of different particles in the ensemble and why contributions are markedly different is a large undertaking and will be addressed in future works.

In field scale, the observations of diverse fault activity is suggested to be explained by a model where plate boundaries are considered as macroscopic granular shear zones near the jamming transition [Ben-Zion, 2008] with effective particle sizes bigger than 10 km enabling earthquakes themselves to redistribute forces within plate boundaries by rearranging force chains [Meroz & Meade, 2017]. Our work shows the potential of numerical methods and in particular, the discrete element approach to improve the understanding of frictional processes that dictate fault frictional slip. This letter also highlights the ability of ML algorithms to characterize the highly irregular frictional behavior of a fault system and identify signals and parameters of importance to the modelling of such systems.

## Acknowledgments

OD and JC thank ETH Zürich for funding and Empa for infrastructural support. CR, BRL, CH, PJ and RG acknowledge institutional support (LDRD) at Los Alamos for funding. The data related to this paper can be obtained by contacting the corresponding authors at domid@ethz.ch or cren@lanl.gov.

**Figures**

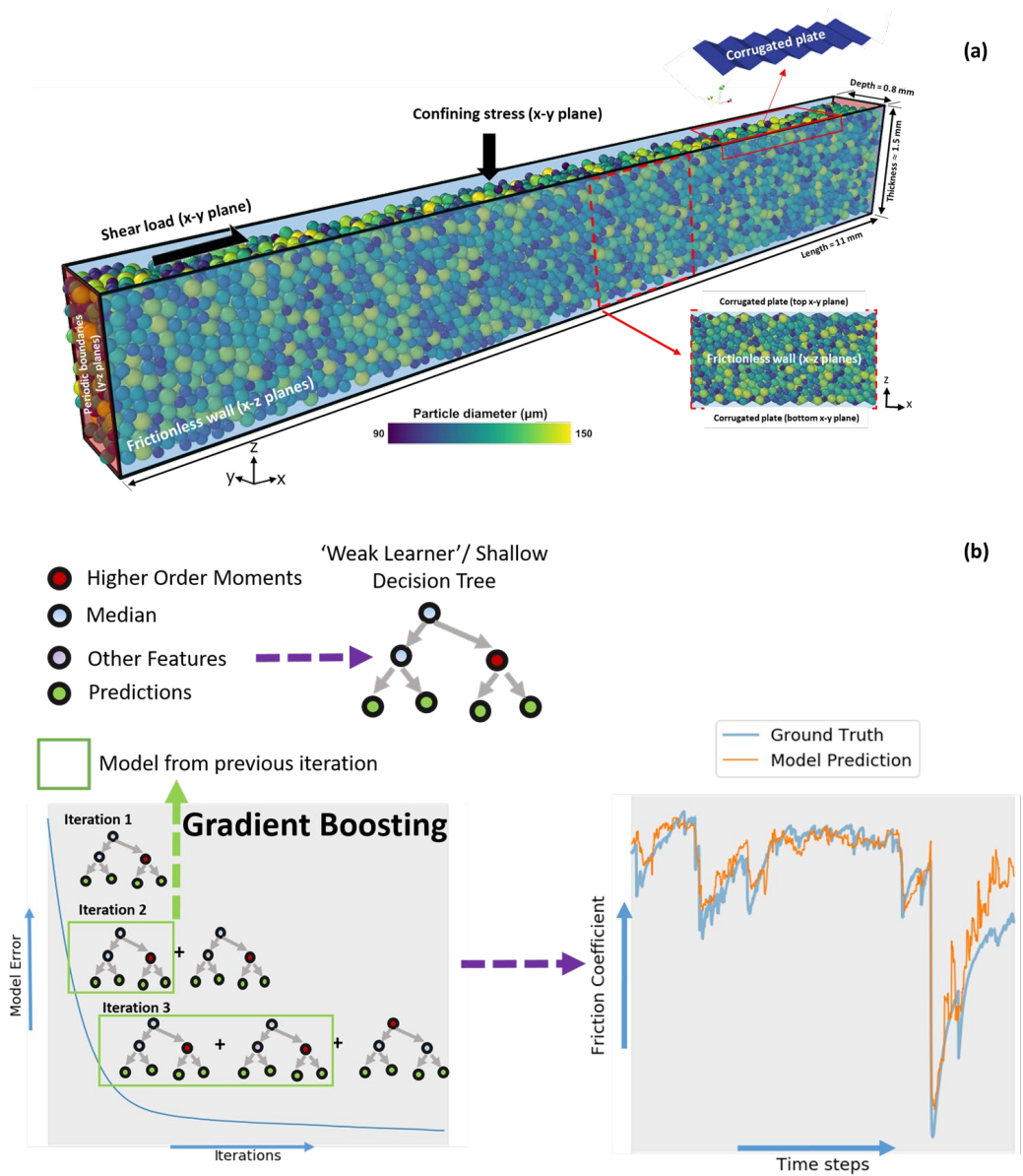

**Figure 1:** (a) Granular fault model with poly-disperse particle diameter distribution of 90-150 micrometer. The fault system is confined in z-direction and sheared in x-direction. Periodic boundary conditions are employed on y-z planes. The shear load is applied along the x-y plane, where two corrugated plates are used on top and bottom x-y planes of the gouge simulating a rough fault surface. (b) A simplified diagram of the gradient boosting process. At each iteration, a weak learner is fit to the residuals from the previous iteration, and is added to the overall model. The model at each iteration is thus an ensemble of the weak learners from previous iterations with an additional weak learner fit to the residuals of the previous ensemble. This iterative process results in the increasing predictive ability of the model, and the reduction of training error.

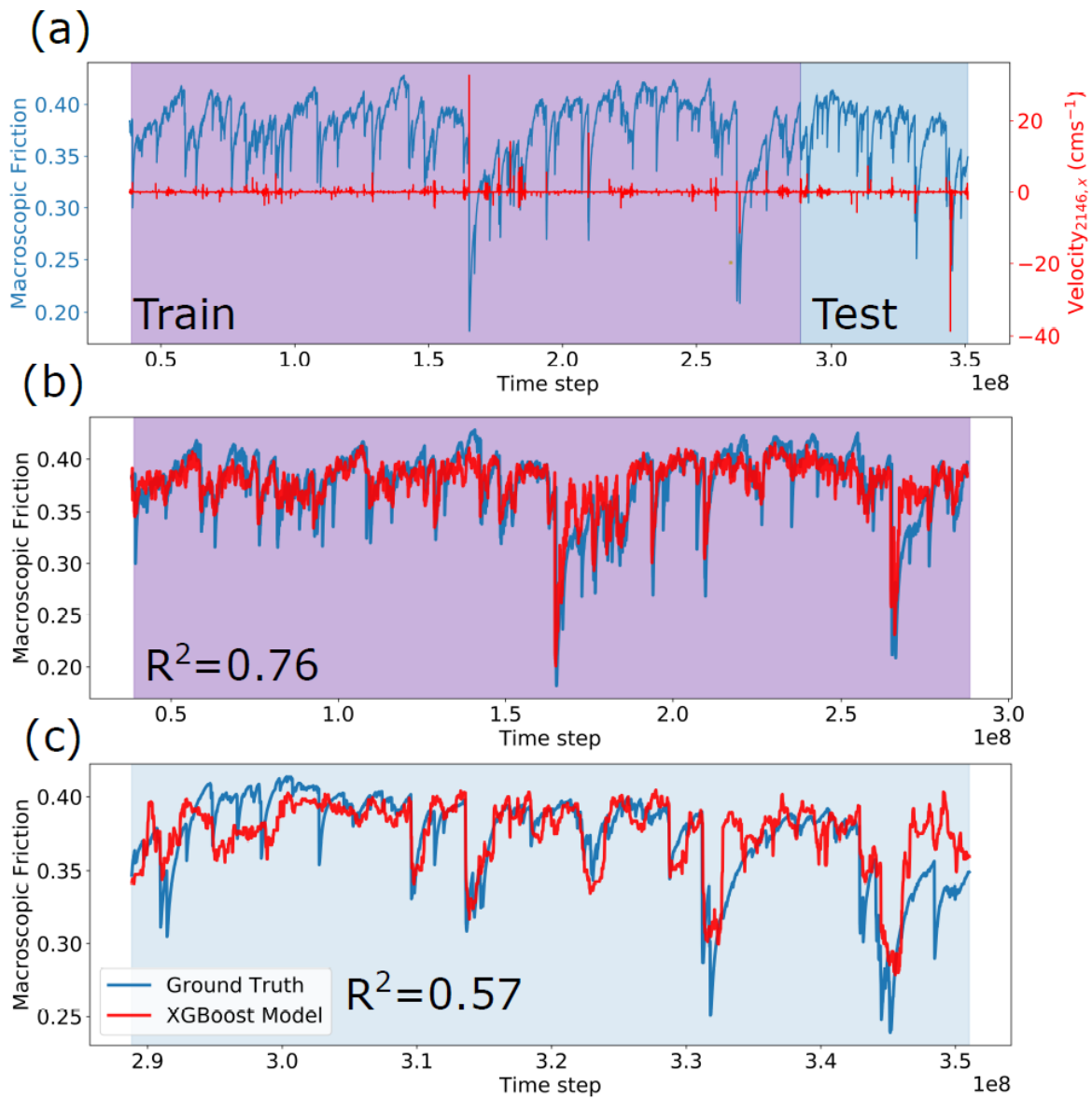

**Figure 2.** XGBoost estimates the instantaneous frictional state of the DEM system from the velocity signal of a single particle. (a) Frictional state of the system (blue curve) throughout the simulation, the violet and blue shaded regions correspond to the training and test labels (target output) respectively. The velocity signal for a single particle is shown in red, this is the raw data stream used to construct features. (b) Performance of the XGBoost model (in red) vs the ground truth (blue) for the training set. (c) Performance of the model for the testing set with a performance of $R^2$=0.57.

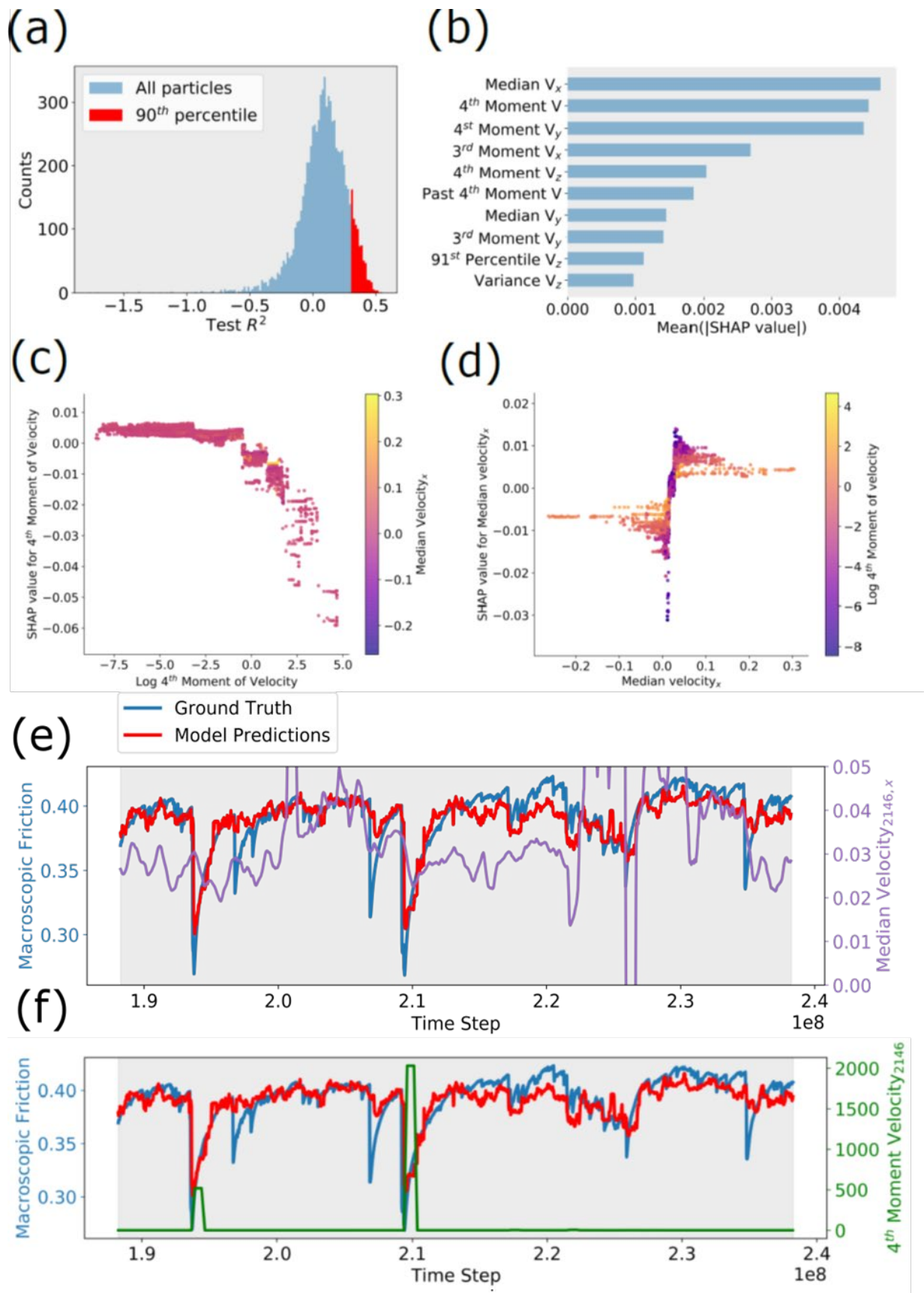

**Figure 3:** Model outputs: (a) Distribution of test $R^2$ scores for XGBoost models trained on features built from the velocity signal of each particle in the simulation, with the 90$^{th}$ percentile of the scores highlighted in red (test $R^2 >=0.3$). (b) Top 10 features for the model shown in Figure 2(b) based on mean absolute SHAP value, ordered from best to worst top down. (c) Dependence plot for the log 4$^{th}$ moment, colored by the median x-component feature. (d) SHAP dependence plot for the median value of the x-component colored by the log 4$^{th}$ moment of the velocity. (e-f) Feature vs model plots for the median x-component and the 4$^{th}$ moment of the velocity signal. The median values have been clipped between 0 and 0.5 for visualization purposes.

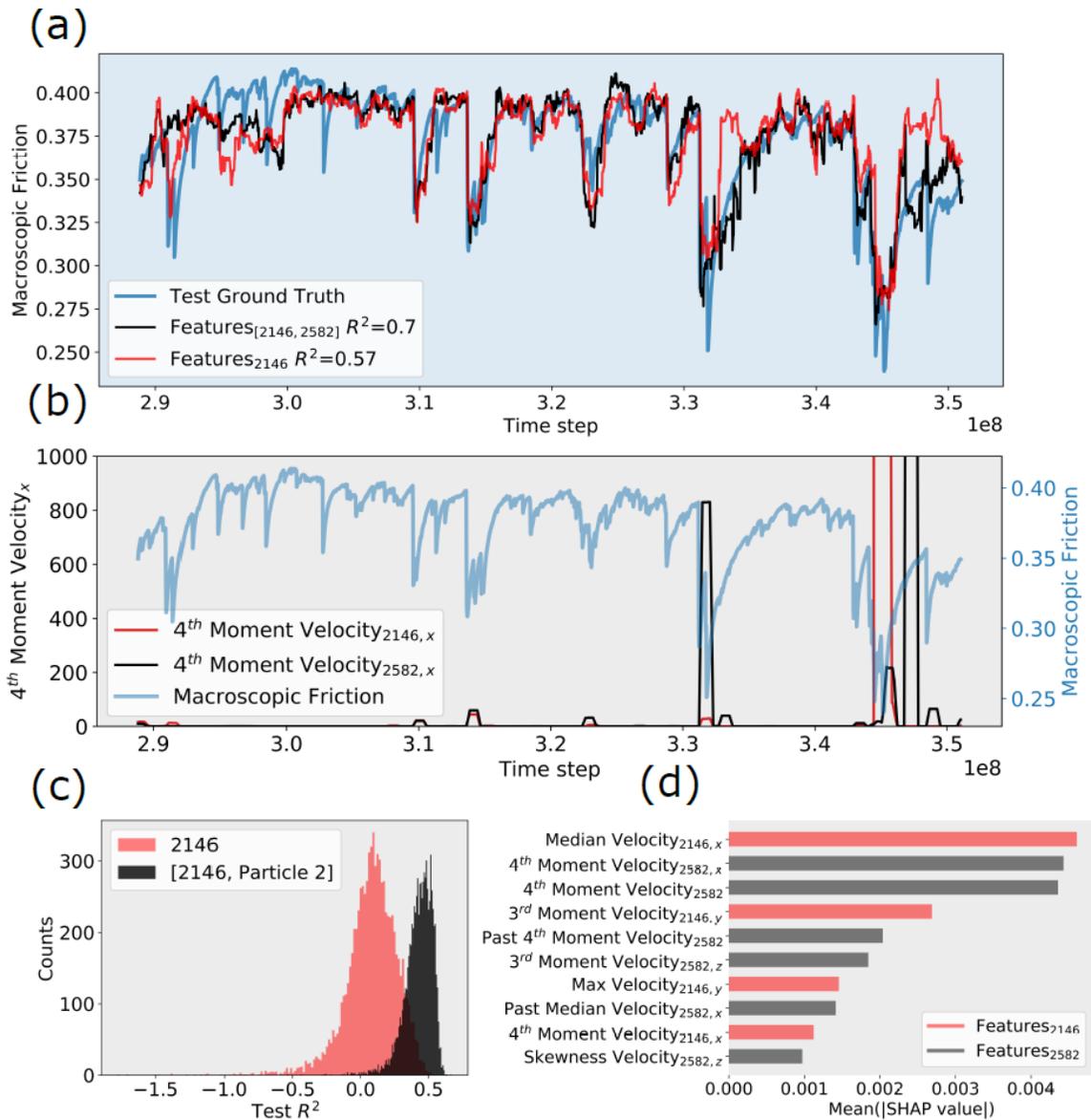

Figure 4: Results of ML model with combined features for two particles: (a) Improved model performance of (black trace, $R^2=0.7$) of XGBoost model trained on features from particles ID 2146 and 2582, compared to using the features only from 2146 (red trace) (b) 4$^{th}$ Moment of the velocity from the two particles, 2146 in red and 2582 in black with the macroscopic friction in blue, we can see that the black trace includes more spikes in the 4$^{th}$ moment. (c) Shift in the distribution of test $R^2$ for single particles (red bins), and all other particles combined with particle 2146 (black bins). (d) Top ten combined features for the model shown in (a).